\documentclass[preprint,showpacs,preprintnumbers,amsmath,amssymb]{revtex4}
\usepackage{dcolumn}
\usepackage{bm}
\usepackage{graphicx}
\usepackage{natbib}

\begin{document}
\bibliographystyle{plainnat}

\preprint{Non-linear effects in ECCD applied for the stabilization of NTMs}

\title{Non-linear effects in electron cyclotron current drive applied for the stabilization of neoclassical tearing modes}

\author{B. Ayten, E. Westerhof}
\email[E-mail: ]{E.Westerhof@differ.nl}
\affiliation{FOM Institute DIFFER, Association EURATOM-FOM, Nieuwegein, Netherlands} \author{ASDEX Upgrade Team} \affiliation{Max Planck Institute for Plasma Physics, EURATOM Association, Garching, Germany}

\date\today

\begin{abstract}
Due to the smallness of the volumes associated with the flux surfaces around the O-point of a magnetic island, the electron cyclotron power density applied inside the island for the stabilization of neoclassical tearing modes (NTMs) can exceed the threshold for non-linear effects as derived previously by Harvey et al, Phys. Rev. Lett. 62 (1989) 426. We study the non-linear electron cyclotron current drive (ECCD) efficiency through bounce-averaged, quasi-linear Fokker-Planck calculations in the magnetic geometry as created by the islands. The calculations are performed for the parameters of a typical NTM stabilization experiment on ASDEX Upgrade. A particular feature of these experiments is that the rays of the EC wave beam propagate tangential to the flux surfaces in the power deposition region. The calculations show significant non-linear effects on the ECCD efficiency, when the ECCD power is increased from its experimental value of 1 MW to a larger value of 4 MW. The nonlinear effects are largest in case of locked islands or when the magnetic island rotation period is longer than the collisional time scale.  The non-linear effects result in an overall reduction of the current drive efficiency for this case with absorption of the EC power on the low field side of the electron cyclotron resonance layer. As a consequence of the non-linear effects, also the stabilizing effect of the ECCD on the island is reduced from linear expectations.
\end{abstract}


\maketitle

\section{Introduction}

Electron cyclotron current drive (ECCD) is used for suppression of neoclassical tearing modes (NTMs) because they limit the performance of tokamak fusion reactors and may lead to plasma disruptions~\cite{lahaye_2006,Kislov,westerhof_NF,Classen,zohm,Gantenbein,Isayama,lahaye_2002,petty,lahaye_2006NF,maraschek,isayama_2009}. A theoretical model for the interpretation of current experiments and extrapolation of their results to future fusion reactors is provided by the generalized Rutherford equation~\cite{rutherford1,rutherford2}. This equation describes the time evolution of the full width $w$ of the magnetic island associated with an NTM. It consists of several terms expressing different stabilizing and destabilizing mechanisms. The term accounting for the contribution of the ECCD to the NTM evolution is denoted as $\Delta^\prime_{\rm ECCD}$. It is proportional to the two dimensional integral over the cross section of the island of that fourier component of the EC driven current density that corresponds to the helicity of the island. In the calculation of $\Delta^\prime_{\rm ECCD}$, the EC power deposition profile and driven current density profile are often averaged over one island rotation period $\tau _{rot}$. The driven current density $j_{\rm ECCD}$ is related to the power deposition $p_{\rm ECCD}$ through the ECCD efficiency $\eta_{\rm ECCD} \equiv j_{\rm ECCD}{\rm d}S/p_{\rm ECCD}{\rm d}V$, which is generally obtained in the linear regime where the ECCD efficiency is power independent. This neglects any temporal variation as a consequence of rotation of the island through the EC power deposition region as well as possible nonlinearities in the ECCD efficiency at high power densities due to modifications of the electron distribution function as a consequence of EC driven quasi-linear velocity space diffusion~\cite{harvey,prater}.

In a previous study we analyzed the consequences of removing the averaging over the island rotation period~\cite{ayten}. It was shown that for rotation periods of the order of, or longer than the electron collision time, the EC driven current density profile moves through the island with the rotation period. Consequently, also $\Delta^\prime_{\rm ECCD}$ exhibits a significant oscillation with the rotation period. This study still relied on the assumption of a linear ECCD efficiency, which implies that for a constant island width the resultant $\Delta^\prime_{\rm ECCD}$ averaged over one rotation period exactly equals the value that is obtained in case of the rotation averaged current density profile. In that case, the oscillations in $\Delta^\prime_{\rm ECCD}$ lead to significant changes of the stabilizing effect of ECCD only when the island width changes significantly during a rotation period.

In the present study we analyze the assumption of a linear ECCD efficiency: under which conditions does the ECCD efficiency become nonlinear and what are the consequences of a non-linear ECCD efficiency. It has been shown by Harvey et al.~\cite{harvey} that non-linear effects appear when the ratio of the absorbed power density over the square of the electron density $n_{e}$ exceeds a certain threshold:
\begin{equation}
\label{eq:harvey}
H \equiv p_{\rm ECCD}\left [ {\rm MW}/{\rm m}^{3} \right ]/\left ( n_{e}\left [ 10^{19}m^{-3} \right ] \right )^{2}\gtrsim 0.5
\end{equation}
where we have introduced the non-linearity parameter $H$. In the non-linear regime the ECCD efficiency is shown to become power dependent: at a given plasma radius the ECCD efficiency increases above linear values for absorption of the EC power on the low field side of the electron cyclotron resonance layer whereas it decreases and even passes through zero for absorption on the high field side of the resonance. In a tokamak the absorbed power density is obtained as a function of the magnetic surfaces. A magnetic island changes the topology of these magnetic surfaces dramatically. Introducing $\Omega$ as the helical flux function which determines the magnetic surfaces in the presence of a magnetic island, $P(\Omega)$ defines the total power absorbed inside the flux surface labeled with $\Omega$. The local power density defined as $p_{\rm ECCD} \equiv {\rm d}P(\Omega)/dV$ is then seen to become large when either ${\rm d}V$ becomes small which is the case for flux surfaces near the O-point of the island \cite{isliker}, or when ${\rm d}P(\Omega)$ itself becomes large. The latter occurs as a consequence of flux expansion in the case of deposition near the X-point of the island. As a result, the non-linear threshold can be exceeded at lower levels of injected power in the presence of a magnetic island than in the unperturbed magnetic equilibrium~\cite{isliker}.

We calculate the non-linear ECCD efficiency through bounce-averaged, quasi-linear Fokker-Planck code calculations in a realistic magnetic geometry with a locked or rotating magnetic island. In this study we restrict ourselves to the case of CW application of the ECCD. The unperturbed plasma equilibrium is taken in accordance with ASDEX Upgrade discharge nr. $26827$ which features a 3/2 NTM~\cite{reich}. While locked 3/2 modes are rare in present day experiments, ECCD control of 2/1 locked modes has been reported in Refs.~\cite{volpe2009,esposito}. This paper is organized as follows: Section $2$ gives the theoretical framework briefly describing the EC ray-tracing and Fokker-Planck codes used. It also includes a description of the flux surfaces in presence of a magnetic island. Power deposition and current drive in the unperturbed equilibrium are analyzed in Section $3$. This serves as a cross code benchmark in the linear regime and provides a basic understanding of the behavior in the nonlinear regime. The results of non-linear effects in the presence of a magnetic island are presented in Section $4$ for the case of a locked island, and in Section $5$ for the case of a rotating island. This is followed in Section $6$ by a study of their consequences on the magnetic island evolution. Finally, a summary and conclusions are provided in Section $7$.

\section{Theoretical framework}

\subsection{ Numerical codes }

In this study we use the TORAY ray tracing code~\cite{Kritz,westerhof_toray} in combination with the RELAX bounce-averaged, quasi-linear Fokker-Planck code~\cite{westerhof_relax}. TORAY applies the cold plasma, Appleton-Hartree dispersion relation to calculate the trajectories of a number of rays. The power absorption along each ray is calculated by the (weakly) relativistic warm plasma dispersion relation. The driven current density is obtained through a linear, adjoint calculation of the current drive efficiency. This employs the subroutines developed by Lin-Liu~\cite{lin-liu} extended with the current response function as derived by Marushchenko~\cite{marushchenko,marushchenko_comment} for a momentum conserving electron-electron collision operator. The information on the ray trajectories is passed to the Fokker-Planck code RELAX. A detailed description of the information transferred from TORAY to RELAX is provided in the Appendix.

The RELAX code solves the bounce-averaged quasi-linear Fokker-Planck equation in toroidal geometry on a finite number of flux surfaces providing an appropriate coverage of the power deposition region. Including only the terms used in the present research this equation is written as~\cite{Killeen}:
\begin{equation}
\label{eq:bounce-averaged quasi-linear Fokker-Planck equation}
\frac{\partial f_{e}}{\partial t}=\left \langle \sum_{s}C\left ( f_{e},f_{s} \right ) \right \rangle_{\phi_{B}}-\left \langle \boldsymbol{\Gamma}_{ql}\right \rangle_{\phi_{B}},
\end{equation}
where $f_{e}$ represents the gyro- and bounce-phase independent electron distribution function. RELAX solves Eq.~\ref{eq:bounce-averaged quasi-linear Fokker-Planck equation} in terms of the low-field-side (lfs) momentum $p_{0}$ and pitch angle $\theta_{0}\left ( = \arccos p_{\parallel 0}/p \right )$, where $\parallel$ represents the component parallel to the magnetic field. The bounce averaging is defined as
\begin{equation}
\left \langle Q \right \rangle_{\phi _{B}}=\frac{1}{\tau_{B}} \oint Q \frac{\mathrm{d} s}{v \cos \theta },
\end{equation}
where $\mathrm{d} s$ is the element of the arc length along the magnetic field line associated with the gyro-center motion and $v$ the particle velocity. The bounce period is given by
\begin{equation}
\tau _{B}= \oint \frac{\mathrm{d} s}{v \cos \theta }.
\end{equation}
The first term on the right-hand side of Eq.~(\ref{eq:bounce-averaged quasi-linear Fokker-Planck equation}), $C(f_{e}, f_{s})$, is the collision term representing the rate of change in the electron momentum distribution as a result of collisions with species $s$. In the present research the electron collision operator is approximated by a truncated linearized collision operator~\cite{Karney}. It is composed of three parts, one representing the effect of the collisions off a background Maxwellian population on the electron distribution function and the other one ensuring the conservation of momentum in electron-electron collisions, that is essential for an accurate calculation of the driven current density. The electron-ion collisions are taken to contribute only to the the pitch angle scattering term in the high velocity limit. They are characterized by an effective charge of the ions, $Z_{\rm eff}$.

The second term on the right-hand side of Eq.~\ref{eq:bounce-averaged quasi-linear Fokker-Planck equation} accounts for the quasi-linear diffusion driven by the EC waves. As the diffusion driven by the EC waves generally occurs in the direction of the perpendicular momentum~\cite{prater}, its corresponding diffusion operator is conveniently written in terms of the diffusion of the invariant magnetic moment $\mu = p_{\perp }^{2}/2m_{e}B$ where $m_{e}$ is the electron mass. The diffusion coefficient of the magnetic moment, $D_{\mu \mu }$ is given by~\cite{westerhof_relax,Peeters}
\begin{equation}
D_{\mu \mu }=\frac{\pi e^{2}}{m_{e}^{2}\omega}\frac{\gamma p_{\perp }^{2}}{B^{2}}\left | \bar{G}_{\perp } \right |^{2}\sqrt{\frac{\pi}{ 2\Delta Q}}e^{-\left ( \gamma -n\omega_c/\omega-N_{\parallel} x_{\parallel} \right )^{2}/\left ( 2\Delta Q \right )}\frac{P_{inj}e^{-\tau}}{\Pi \cos \chi } \frac{B}{2\pi \tau_{B} v_{\parallel}R B_{p}},
\end{equation}
where $\omega$ is the EC wave frequency, $\omega _{c}$ the electron cyclotron frequency, $n= 1,2,...$ the harmonic number, $\gamma$ the relativistic factor, $N_{\parallel}$ the parallel refractive index, $x_{\parallel}=p_{\parallel}/m_{e} c$ the normalized parallel momentum, $v_{\parallel}$ the parallel velocity of the particle, $R$ the major radius, and $B_{p}$ the poloidal component of the magnetic field. The factor $\left | \bar{G}_{\perp}\right |^{2}$, where
\begin{equation}
\bar{G}_{\perp} = \frac{n \omega_{c}}{v_{\perp}\gamma \omega}\left [ v_{\perp} \left ( \varepsilon^{+}J_{n+1}\left ( b \right ) + \varepsilon^{-}J_{n-1}\left ( b \right ) \right )+v_{\parallel}\varepsilon_{\parallel}J_{n}\left ( b \right )\right ],
\end{equation}
accounts for the effect of wave polarization. The wave polarization vector is normalized such that its amplitude $\varepsilon = 1$ and $\varepsilon^{\pm } = \varepsilon_{x} \pm i\varepsilon_{y}$ in a frame where the magnetic field is along the $z$-coordinate and the perpendicular wave vector along the $x$-coordinate. $J_{n}$ is the Bessel function of the first kind of (integer) order $n$ and its argument $b\equiv k_{\perp }\rho_{L}$ where $k_{\perp }$ represents the magnitude of the perpendicular wave vector and $\rho_{L}$ the Larmor radius of the electrons. The next term defines the broadened resonance condition with the total resonance broadening $\Delta Q$ given by
\begin{equation}\label{eq:broadening}
\Delta Q=\left [ \frac{L_{\vartheta }\partial}{r \partial \vartheta}\left ( \gamma-n\omega_{c}/\omega-N_{\parallel}x_{\parallel} \right ) \right ]^{2} + \left [ \frac{L_{\varphi }\partial}{r \partial \varphi}\left ( \gamma-n\omega_{c}/\omega-N_{\parallel}x_{\parallel} \right ) \right ]^{2} + \frac{\gamma^{2}R^{2}\dot{\varphi }^{2}}{\omega L_{\varphi }^{2}},
\end{equation}
where $R \dot{\varphi }$ refers to the toroidal velocity. The first two terms denote the broadening as a consequence of the variation in magnetic field and parallel refractive index over the beam in poloidal and toroidal directions, respectively. The beam power profile is assumed to be Gaussian in both the poloidal and toroidal directions with widths of $L_{\vartheta }$ and $L_{\varphi }$, respectively. The first two terms of Eq.~$\ref{eq:broadening}$ can be obtained from a bounce-averaging of the local delta function resonance~\cite{giruzzi}. The last term describes the resonance broadening owing to the finite wave-particle interaction time during a beam crossing.
The term $P_{inj}e^{-\tau}$ accounts for the total wave power crossing the flux surface weighted by the factor $1 / \Pi \cos \chi $, where $\Pi$ is the power flux for a normalized electric field vector and $\chi$ the angle between the direction of wave propagation and the normal of the flux surface. The factor $e^{-\tau}$ where $\tau$ is the optical depth $\tau = 2\int k_{i} \mathrm{d} s$ ($k_{i}$ represents the imaginary part of wave vector), refers to the power fraction absorbed so far along the beam trajectory $s$. The term $B/\left ( 2\pi \tau_{B} v_{\parallel}R B_{p} \right )$ is a division by the effective flux surface area.

\subsection{ Plasma equilibrium and magnetic topology }

\begin{figure}[!hbtp]
\includegraphics[width=80mm]{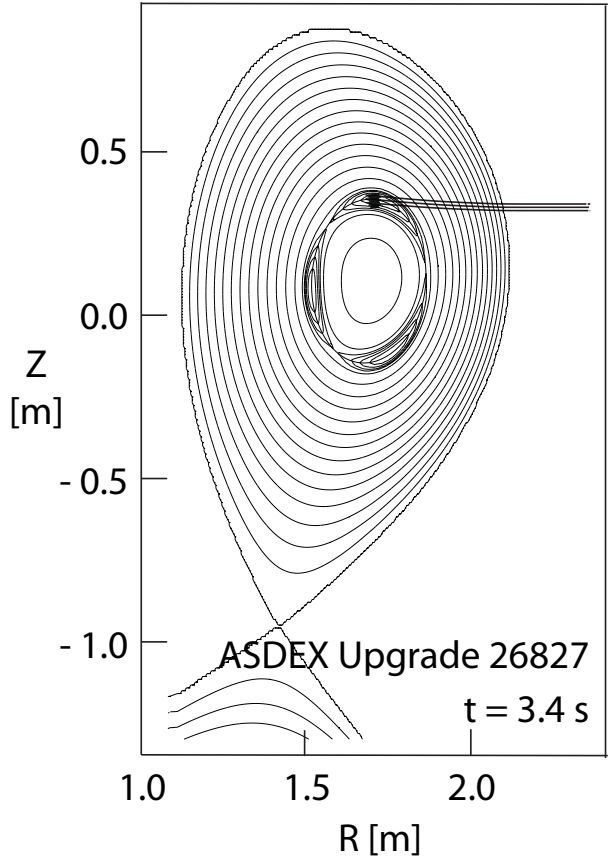}
\caption{\label{figure1}ASDEX Upgrade equilibrium for discharge nr.~26827 with a 4 cm 3/2 magnetic island superimposed. The island phase $\xi_0 =-0.5$ such that the ECCD deposition is near the O-point of the magnetic island. Also shown are the trajectories of a number of rays modeling the injected ECCD beam (the injection parameters are specified in Section 3). The poloidal cross section of the helical flux surfaces is displayed for the toroidal position of ECCD power deposition.}
\end{figure}

\begin{figure}[!hbtp]
\includegraphics[width=80mm]{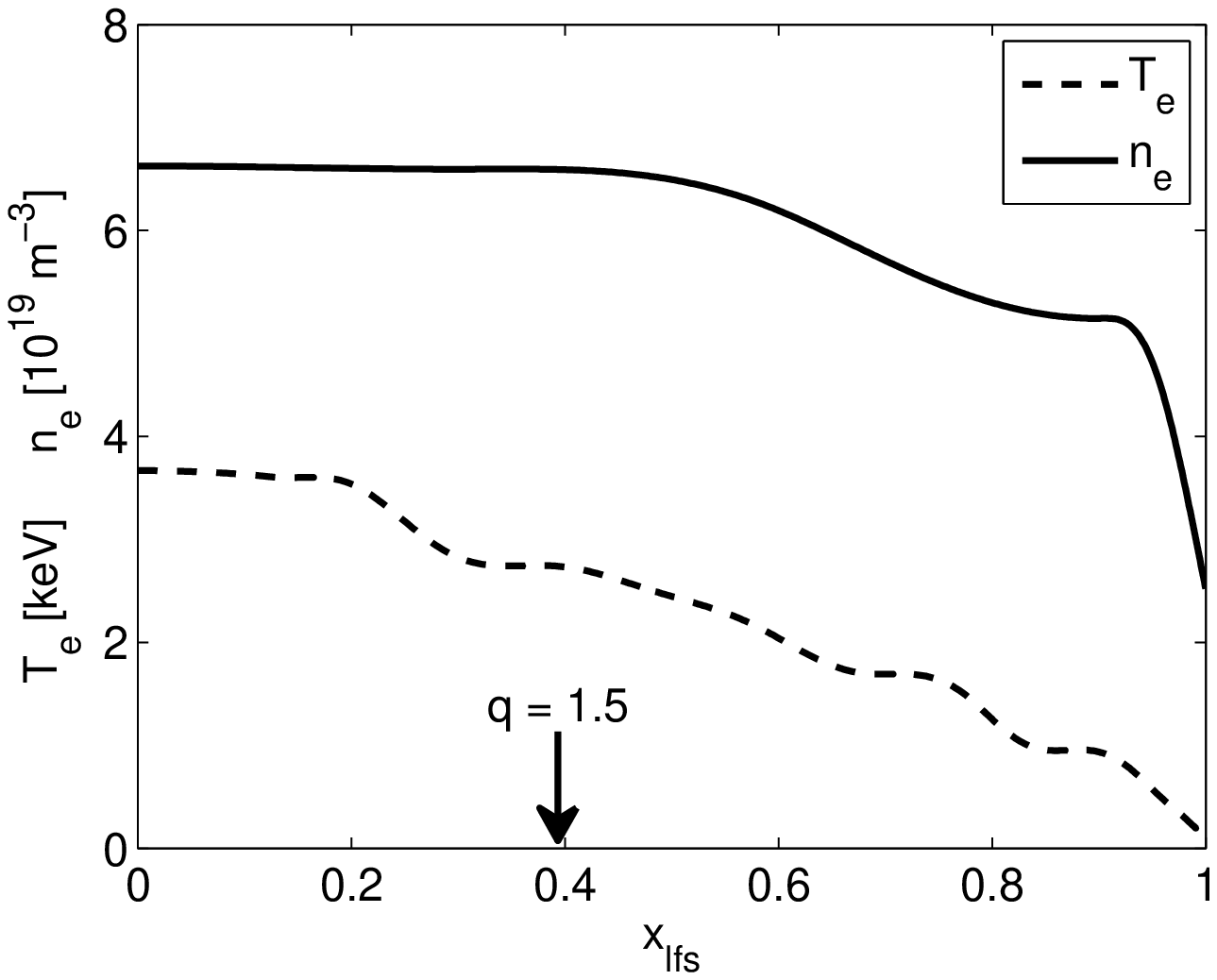}
\caption{\label{figure2}The density and temperature profiles at $t=3.4$~s obtained from the IDA Integrated Data Analysis diagnostic of ASDEX Upgrade. Note that these profiles are shown as a function of the normalized minor radius in the low field side mid plane, $x_{\rm lfs}$. The position of the $q=m/n=3/2$ resonant surface is indicated.}
\end{figure}

As the basis of our study we use a typical discharge from ASDEX Upgrade in which ECCD has been used to suppress an NTM with poloidal and toroidal mode numbers $m=3$, and $n=2$, respectively~\cite{reich}. Figure~\ref{figure1} shows the experimental equilibrium for ASDEX Upgrade discharge nr.~26827 ($B_{t} = 2.6$~T, $I_{p} = 1$~MA, $T_{e}\left ( 0 \right ) = 3.7$~keV, $n_{e}\left ( 0 \right ) = 6.6\times 10^{19}$~m$^{-3}$) at $t=3.4$~s around the time that the ECCD is applied. Figure~\ref{figure2} shows the density and temperature profiles at this time as obtained from the IDA Integrated Data Analysis diagnostic of ASDEX Upgrade. These profiles are shown as a function of the normalized minor radius in the low field side mid plane, $x_{\rm lfs}$. For the calculations of the current drive efficiency in both TORAY and RELAX, the effective charge of the ions has been taken to be $Z_{\rm eff} = 1.6$.

NTMs perturb the equilibrium magnetic field topology: in the case of a single helicity perturbation with poloidal mode number $m$ and toroidal mode number $n$, the unperturbed equilibrium topology of closed nested surfaces of constant poloidal flux is replaced by a topology including a single chain of magnetic islands around the surface with the resonant safety factor $q=m/n$. This perturbed flux surface topology is represented by the surfaces of constant helical flux. The unperturbed helical flux function $\psi_0$ is obtained from the experimental equilibrium. The topology in the presence of a magnetic island is then obtained by superimposing a helical flux perturbation $\tilde{\psi}$ as~\cite{Yang}
\begin{equation}\label{eq:helical flux perturbation}
\tilde{\psi} = \tilde{\psi }\left ( r_{c,s} \right )\frac{r_{c}^{2}}{r_{c,s}^{2}}\frac{\left ( 1-\frac{r_{c}}{a} \right )^{2}}{\left ( 1-\frac{r_{c,s}}{a} \right )^{2}} \cos \left ( m\theta _{s} + n\phi   + \xi _{0}\right )
\end{equation}
which provides a realistic approximation to the radial dependence of the tearing perturbation. In particular, this equation describes a magnetic island which has a significantly asymmetric width on opposite sides of the resonant surface $r_{c,s}$, consistent with experimental observations. We have introduced here the coordinate $r_c$ as a flux surface label of the unperturbed equilibrium: $r_{c} \equiv \sqrt{S\left ( \psi _{0}, \sigma  \right )/\pi}$, where $S$ refers to the surface area enclosed by the surface with helical flux $\psi _{0}$ and $\sigma$ indicates the position relative to the resonant surface $r_{c,s}$ with $\sigma=-1$ for $r_{c} < r_{c,s}$ and $\sigma=+1$ for $r_{c} > r_{c,s}$. In Eq.~\ref{eq:helical flux perturbation}, $a$ is the value of $r_{c}$ at the edge of the plasma, $\theta _{s}$ the straight field line poloidal angle, $\phi$ the toroidal angle, and $\xi_{0}$ the phase of the island with $\xi_{0}=0$ meaning that the X-point is on the lfs mid-plane at $\phi=0$. The perturbation amplitude at the resonant surface $\tilde{\psi }\left ( r_{c,s} \right )$ is adjusted to obtain a specific island size. Throughout the paper island sizes will be given in terms of their full width (i.e. the maximum distance between their separatrices) as measured on the low field side mid plane. Note that the origin of the toroidal angle is chosen to coincide with the position of the ECCD launching mirrors. As a result deposition on the O-point of the island is obtained in case of an island phase $\xi_0 = -0.5$ (see Fig.~(\ref{figure1})).

The perturbation in the equilibrium temperature and density profiles as a result of the presence of an island are taken into account in TORAY. It is noted, however, that over the parameter range studied their effect on the wave propagation was very small. We assume that temperature and density are constant on the magnetic flux surfaces defined by the perturbed helical flux $\psi = \psi_{0}+\tilde{\psi }$ and $\sigma$. The radial transport outside the island is assumed to be unaffected, therefore we relate the temperature $T\left ( r_{c},\theta,\phi\right )$ at $r_{c} > r_{c,s}$ outside the island to the unperturbed equilibrium temperature $T_{0}\left ( r_{c} \right)$ as\cite{hugovdbrand}:

at position $\left ( r_{c},\theta,\phi\right )$ where $\psi \left ( r_{c}> r_{c,s},\theta,\phi\right )> \psi _{sep}$ and $\sigma = +1$
\begin{equation}
T\left ( r_{c},\theta,\phi\right )=T_{0}\left ( \sqrt{\frac{S\left ( \psi \left ( r_{c},\theta ,\phi  \right ),\sigma  \right )}{\pi }} \right ),
\end{equation}
where $\psi _{sep}$ is the helical flux at the separatrices and $S\left ( \psi ,\sigma\right )$ the area in the poloidal cross section that is enclosed by the flux surface defined by $\psi$ and $\sigma$. The temperature inside the island is assumed constant owing to fast parallel transport and taken to be equal to that of the outer separatrix:

at position $\left ( r_{c},\theta,\phi\right )$ where $\psi \left ( r_{c},\theta,\phi\right )\le \psi _{sep}$
\begin{equation}
T_{isl}=T_{0}\left ( \sqrt{\frac{S\left ( \psi _{sep},+1 \right )}{\pi }} \right ).
\end{equation}
The temperature at $r_{c} < r_{c,s}$ outside the island decreases by an amount $\Delta T$ with regard to the unperturbed equilibrium temperature profile where $\Delta T$ is the difference of the temperatures in the unperturbed equilibrium temperature profile on those surfaces enclosing the same surface area as the inner and outer separatrices, respectively:

at position $\left ( r_{c},\theta,\phi\right )$ where $\psi \left ( r_{c}< r_{c,s},\theta,\phi\right )> \psi _{sep}$ and $\sigma = -1$
\begin{equation}
T\left ( r_{c},\theta,\phi\right )=T_{0}\left ( \sqrt{\frac{S\left ( \psi \left ( r_{c},\theta ,\phi  \right ),\sigma  \right )}{\pi }} \right ) - \Delta T,
\end{equation}

\begin{equation}
\Delta T= T_{0}\left ( \sqrt{\frac{S\left ( \psi _{sep},-1 \right )}{\pi }} \right )-T_{0}\left ( \sqrt{\frac{S\left ( \psi _{sep},+1 \right )}{\pi }} \right ).
\end{equation}
The perturbation in the density profile is obtained analogously. We find that over an extended region around the island, the density and temperature are almost constant and equal to $6.56\times 10^{19}$~m$^{-3}$and $2.7$~keV, respectively. These values are used for their corresponding parameters in the RELAX calculations. The magnetic field perturbation arising from the presence of an island is very small compared to $B_{\phi}$, and is not taken into account in TORAY or RELAX. In particular, for RELAX this means that the bounce integrals representing the effects of trapped particles are evaluated as for the unperturbed equilibrium. The full perturbed flux surface topology, however, is taken into account in the evaluation of the EC quasi-linear diffusion operator. Power deposition and driven current density profiles in both TORAY and RELAX are evaluated in the helical flux surface coordinates of the perturbed equilibrium. These profiles will be reported below in terms of their projection along the low field side mid plane in the toroidal cross section where the island O-point is in the lfs mid plane.

\section{ECCD in the unperturbed equilibrium}

\begin{figure}[!hbtp]
\includegraphics[width=80mm]{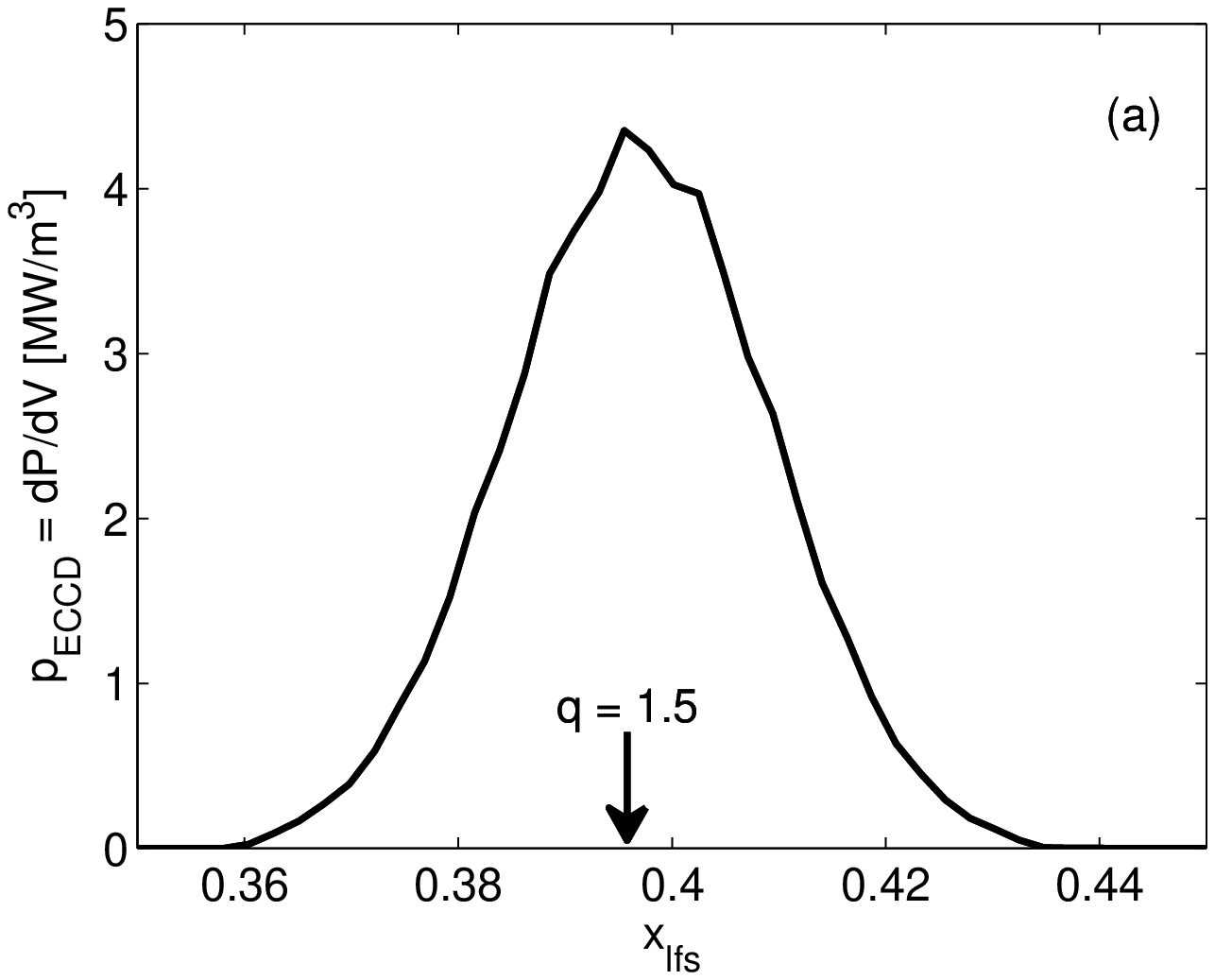}
\includegraphics[width=80mm]{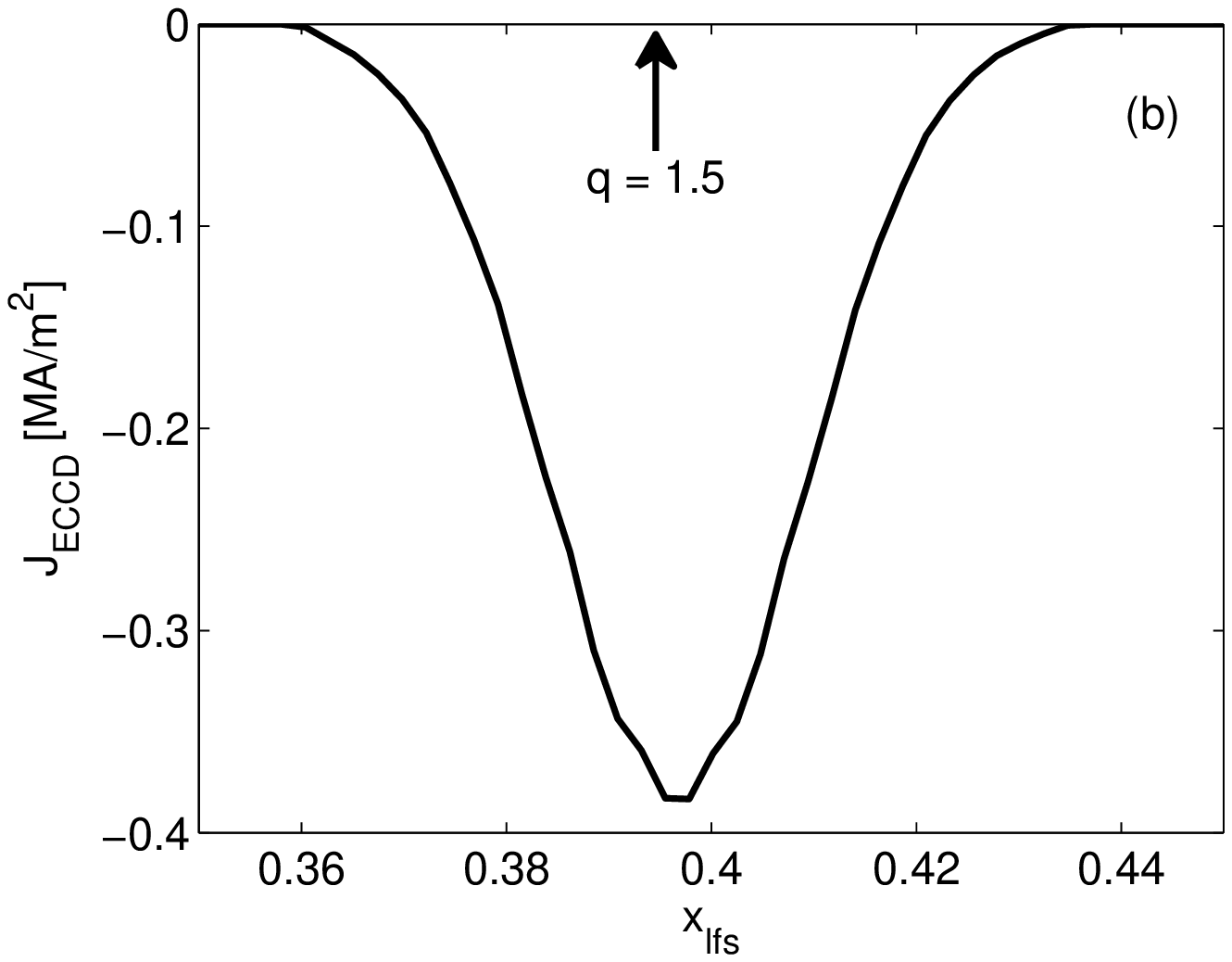}
\caption{\label{figure3}The power deposition (a) and current density profiles (b) calculated with TORAY for the injection of a $1$~MW ECCD beam. Both profiles are centered on the $q=3/2$ surface and are highly localized. The power deposition profile in the unperturbed equilibrium has a maximum corresponding to a value of the non-linearity parameter of $H_{\rm max} = 0.1 \left [ {\rm MW}/{\rm m}^{3} \right ]/\left [ 10^{19}{\rm m}^{-3} \right ]^2$ for these parameters in ASDEX Upgrade. The negative current density (i.e. in the clockwise direction when viewing the tokamak from above) is in the same direction as the Ohmic plasma current.}
\end{figure}

We first study ECCD in the unperturbed magnetic equilibrium. In the example studied we assume that ECCD at $140$~GHz is injected from a top mirror of ASDEX Upgrade with a toroidal injection angle of $\phi = -8^{\rm o}$. The ECRH beam, which is focused near the region of power deposition, is modeled in the TORAY code by an appropriate set of rays with a FWHM of 1.7~cm in the vertical direction and FWHM of $3^{\rm o}$ in the toroidal injection angle. The beam is represented by a square grid of rays with $41$ equidistant points in the poloidal direction and $9$ points in the toroidal direction covering a range of twice the Gaussian width of the beam power density. Each ray is assigned a power fraction in accordance with the Gaussian profile of the beam power density. Note that the rays propagate tangential to the flux surfaces in the power deposition region. As a result, the whole power from a single ray is absorbed over a very narrow range of flux surfaces. The power deposition and current density profiles as obtained from TORAY for the injection of a $1$~MW ECCD beam are shown in Fig.~\ref{figure3}a and b, respectively. Both the power deposition and driven current density profiles are centered on the $q=3/2$ surface and are highly localized. The maximum absorbed power density corresponds to a nonlinearity parameter with a value of $H_{\rm max} = 0.1 \left [ {\rm MW}/{\rm m}^{3} \right ]/\left [ 10^{19}{\rm m}^{-3} \right ]^2$, indicating that nonlinear effects are not expected for these parameters in ASDEX Upgrade in the absence of magnetic islands.

\begin{figure}[!hbtp]
\includegraphics[width=80mm]{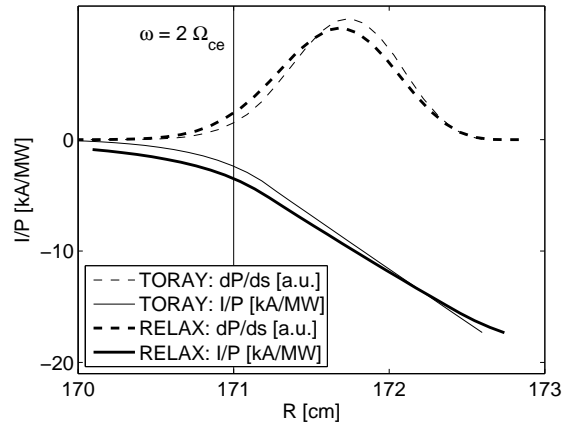}
\caption{\label{figure4}Power absorption ($\rm{d} P/\rm{d} s$ [a.u.], dashed curves) and ECCD efficiency ($\rm{d} I/\rm{d} P$ [kA/MW], solid curves) along the central ray of a low power ECCD beam as a function of the major radius along the ray trajectory calculated with either TORAY (thin curves) or RELAX (thick curves) in the unperturbed magnetic equilibrium. The cold second harmonic EC resonance is at $R = 171$~cm.}
\end{figure}

To obtain a better understanding of the absorption and current drive efficiency, we look into detail at the power absorption along a single ray in both TORAY and RELAX. The chosen ray corresponds to the central ray of the full beam. Because of the tangent propagation of the ray in the region of power deposition this also involves just a single discrete volume shell in RELAX. Figure~\ref{figure4} presents the power absorption ($\rm{d} P/\rm{d} s$ [a.u.]) and ECCD efficiency ($\rm{d} I/\rm{d} P$ [kA/MW]) along the central ray calculated with either TORAY (from a linear, adjoint calculation of the current drive efficiency accounting for momentum conservation in electron-electron collisions~\cite{lin-liu,marushchenko,marushchenko_comment}) or RELAX in the case of low EC power. The results show good agrement and thus, provide a cross code benchmark between TORAY and RELAX in the low power, linear regime.

\begin{figure}[!hbtp]
\includegraphics[width=80mm]{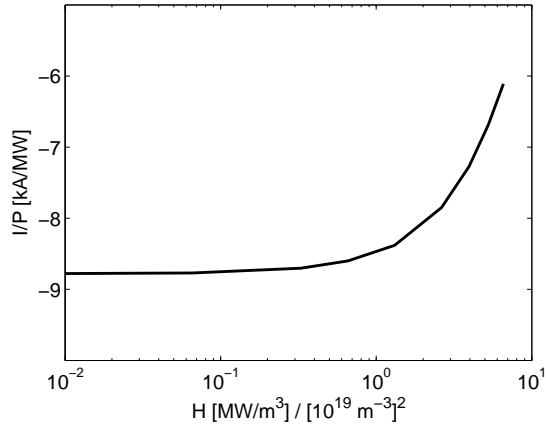}
\caption{\label{figure5}RELAX results for the ECCD efficiency ($\rm I/\rm P$ [kA/MW]) along the central ray of an ECCD beam as a function of its corresponding non-linearity parameter $H$ calculated for a range of different injected power levels. The predicted threshold for nonlinear effects is at $H = 0.5 \left [ {\rm MW}/{\rm m}^{3} \right ]/\left [ 10^{19}{\rm m}^{-3} \right ]^2$~\cite{harvey}. Typical values that could be reached in the experiment range from $H_{\rm max} = 0.1$ for the injection of 1~MW in the unperturbed equilibrium up to $H_{\rm max} = 5 \left [ {\rm MW}/{\rm m}^{3} \right ]/\left [ 10^{19}{\rm m}^{-3} \right ]^2$ for injection of 4~MW localized near the O-point of a magnetic island.}
\end{figure}

In order to study the nonlinear effects of ECCD in this case, where the ray propagates tangentially to the flux surface in the power deposition region, the RELAX calculation is repeated for increasingly higher powers injected along the ray resulting in increasing values of the nonlinearity parameter $H$. In this non-linear regime it is no longer possible to single out the contribution from each individual segment, $\rm {d} s$, of the ray to the driven current ($\rm{d} I/\rm{d} s$). The results show a nonlinear reduction in the absolute value of the integrated ECCD efficiency. Here, the integrated ECCD efficiency is defined as the ratio of total driven current in the volume shell to the total absorbed power in that shell. This is demonstrated in Fig.~\ref{figure5}, which shows the results of the single ray, single volume shell RELAX calculations as a function of the nonlinearity parameter $H$. Note that the range in $H$ covered would include unrealistically high values of the injected power for the case of the unperturbed equilibrium. In the presence of an island, however, a value of $H \approx 5 \left [ {\rm MW}/{\rm m}^{3} \right ]/\left [ 10^{19}{\rm m}^{-3} \right ]^2$ might well be reached with the injection of about 4 MW of ECCD. Significantly higher values of $H$ would also apply in case of experiments at a lower density. The threshold for the non-linear effects where the ECCD efficiency diverts from being a constant and starts decreasing as $H$ increases, is observed to be around $H = 0.5 \left [ {\rm MW}/{\rm m}^{3} \right ]/\left [ 10^{19}{\rm m}^{-3} \right ]^2$ in accordance with the conclusions of Harvey et al.~\cite{harvey}. However, the decrease of the global current drive efficiency is in apparent contradiction with the results of Harvey et al.~\cite{harvey} which for the present conditions (absorption on the low field side of the resonance) would predict an increase in the local current drive efficiency. Note however that Harvey et al. calculated a local current drive efficiency that applies to a given point along a ray, whereas the present calculation is global and thus averages over the full absorption profile. The study of Harvey et al. also extended too much higher power density levels than the current work.

\begin{figure}[!hbtp]
\includegraphics[width=80mm]{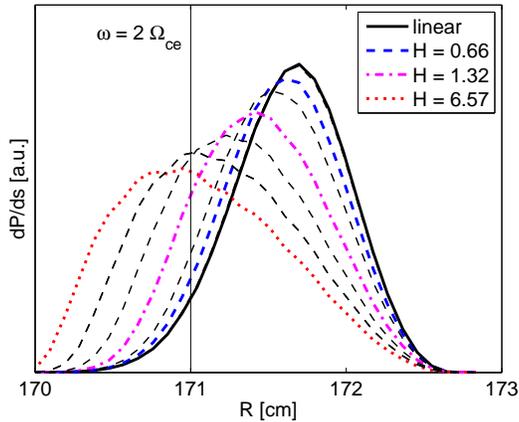}
\caption{\label{figure6}Power absorption normalized to the total injected power ($\rm{d} \bar{P}/\rm{d} s$ [a.u.]) along the central ray of a low power ECCD beam as a function of major radius R calculated with RELAX for different values of non-linearity parameter $H$. The solid black line represents the low power, linear regime, and the dashed curves represent results with increasing power and non-linearity parameter $H$. With the maxima moving to the left, the curves are obtained for $H= 0.66$, $1.32$, $2.63$, $3.94$, $5.26$, and $6.57 \left [ {\rm MW}/{\rm m}^{3} \right ]/\left [ 10^{19}{\rm m}^{-3} \right ]^2$, respectively. A subset of these is indicated in the color legend. The cold second harmonic EC resonance is at $R = 171$~cm.}
\end{figure}

In order to explain the global decrease in the absolute value of the current drive efficiency, Figure~\ref{figure6} shows the power absorption normalized to the injected power ${\rm d} \bar{P}/{\rm d}s$ (with $\bar P \equiv P / P_{\rm injected}$) along the ray for several values of the nonlinearity parameter $H$. At higher power densities, and consequently, at higher values of $H$ the EC driven quasi-linear velocity space diffusion alters the electron distribution function. The quasi-linear flattening of the distribution function at the resonance reduces the absorption coefficient and shifts the peak in the $\rm{d} \bar{P}/\rm{d} s$ profile to a smaller major radius R~\cite{peeters1996} coming closer to the cold second harmonic EC resonance at $R = 171$~cm. That broadens the power deposition profile along the ray with more power reaching the regions where the ECCD efficiency is lower. As a consequence, the integrated ECCD efficiency decreases. We note here that this result of a globally decreasing current drive efficiency is not specific to the present case with tangent ray propagation, but is also obtained when the beam is injected in the mid plane and the rays are propagating close to transverse to the flux surfaces. In that case, the shift in the power deposition along the ray also results in a shift in power deposition in terms of the flux surfaces. Locally on a flux surface the nonlinear current drive efficiency then shows the expected increase consistent with the results of Harvey et al.~\cite{harvey}. Globally, however this local increase in the efficiency is more than compensated by the lower efficiency of the wave power reaching further into the plasma and coming much closer to the cold EC resonance.

\section{ECCD in the presence of a locked magnetic island}
\begin{figure}[!hbtp]
\includegraphics[width=80mm]{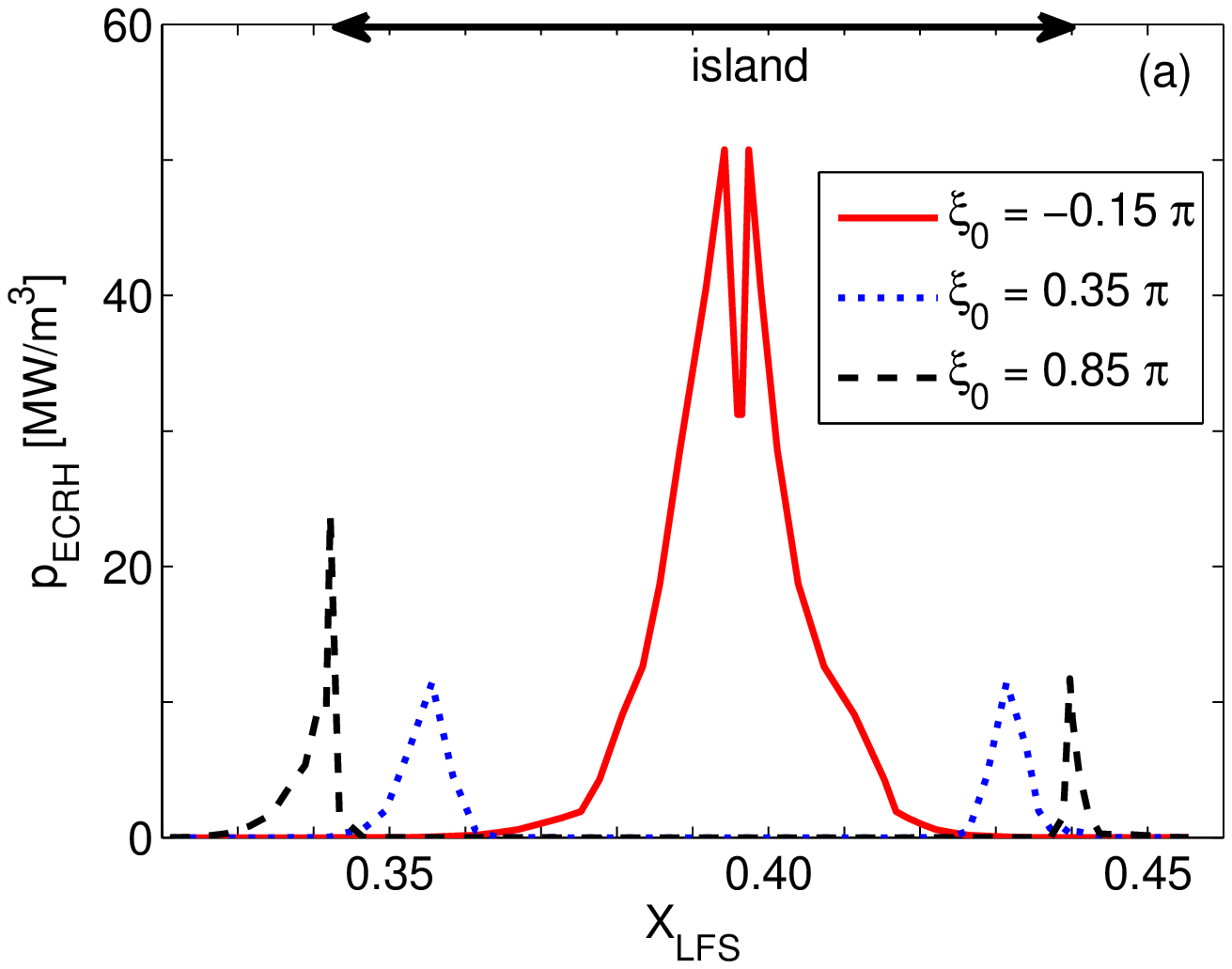}
\includegraphics[width=80mm]{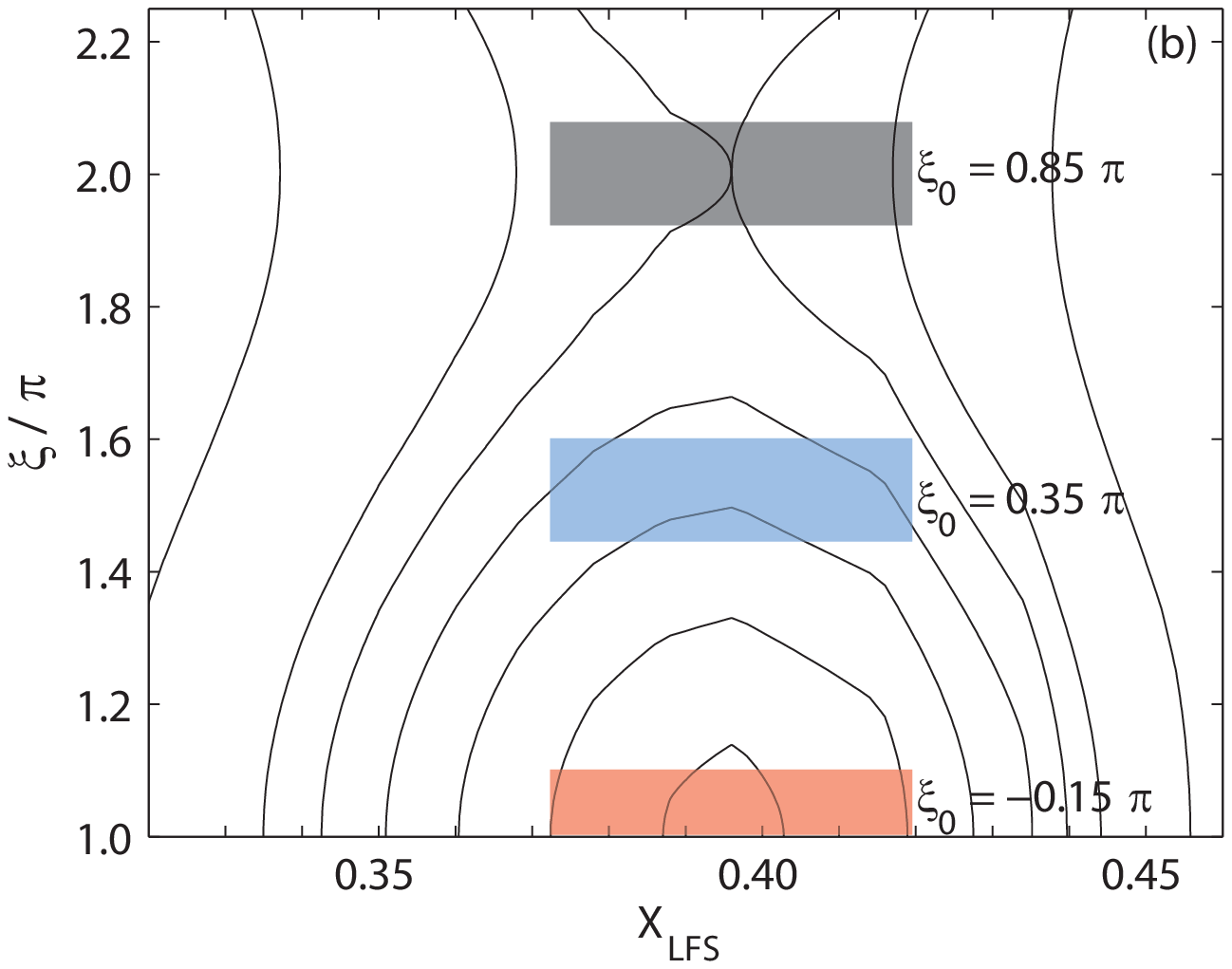}
\caption{\label{figure7}(a) Absorbed power profiles calculated with RELAX for power depositions at three different phases of a locked island. A phase of $\xi_0 = -0.15 \pi$ results in power deposition near the O-point, while $\xi_0 = 0.85$ corresponds to power deposition near the X-point. (b) A sketch of the corresponding power deposition areas projected on the island in coordinates $x_{\rm lfs}$ and $\xi=m\theta+n\phi+\xi_0$ (see text). The black solid lines represent the flux surfaces.}
\end{figure}

We now study ECCD in the presence of locked islands of different sizes. Figure~\ref{figure7} illustrates the power deposition profiles for a $1$ MW ECCD beam in the case of a $4$ cm wide island. The power deposition profile is obtained as a function of helical flux. It is plotted in terms of the normalized low field side minor radius $x_{\rm lfs}$ crossed by the helical flux surface in the poloidal cross section of the plasma in which the O-point of the island is in the low field side mid-plane. The power densities inside the magnetic island are thus plotted twice: for $x_{\rm lfs} < x_{\rm O-point}$ and for  $x_{\rm lfs} > x_{\rm O-point}$. The spatial asymmetry with respect to $x_{\rm O-point}$ in this part of the profile reflects the asymmetric width of the magnetic island. Also note that the deposition profile in general has a discontinuity at the separatrix. The three different profiles correspond to different phases of the island: a phase of $ -0.15 \pi$ corresponding to heating near the O-point, an intermediary case with a phase of $0.35 \pi$, and a phase of $0.85 \pi$ corresponding to heating near the X-point. Here, the phase refers to the angle $\xi_ {0}$ as defined in Section II.B. The peak power density and, consequently, the likelihood of non-linear effects is strongly dependent on the phase of the island. The dip at the center of the deposition profile for heating near the O-point is ascribed to an artifact of the beam discretization: the central volume of the island is very small and is crossed by a limited number of the rays that model the ECCD beam.

\begin{figure}[!hbtp]
\includegraphics[width=80mm]{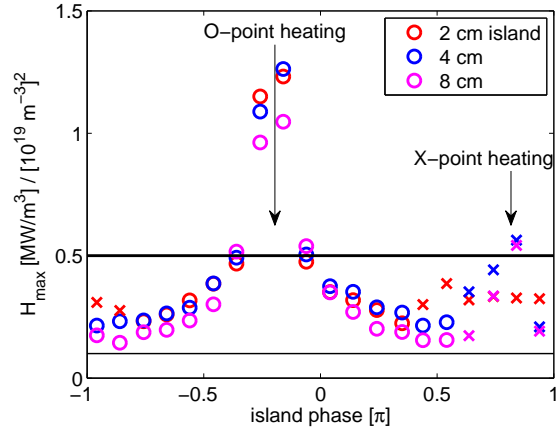}
\caption{\label{figure8}$H_{\rm max}$ as a function of island phase for $w_{island} = 2$, $4$, and $8$~cm. The symbols indicate the position where this maximum is attained: circles for inside the island, and crosses for on the separatrix. The horizontal line indicates the threshold for nonlinear effects, Eq.~(\ref{eq:harvey}).}
\end{figure}

In Fig.~\ref{figure8} we present the results of RELAX calculations for the injection of a $1$ MW ECCD beam in a geometry including a locked island of $2$, $4$ and $8$~cm width for different island phases. The results are given in terms of the peak value $H_{\rm max}$ of the non-linearity parameter (\ref{eq:harvey}). The thin solid black line represents the value of $H_{\rm max}$ in the unperturbed equilibrium, while the thick black line indicates the threshold for non-linear effects. Circles represent cases in which the peak value is attained inside the island and crosses cases in which the peak value is attained on the separatrix. In case of deposition around the O-point the threshold is clearly exceeded.

\begin{figure}[!hbtp]
\includegraphics[width=80mm]{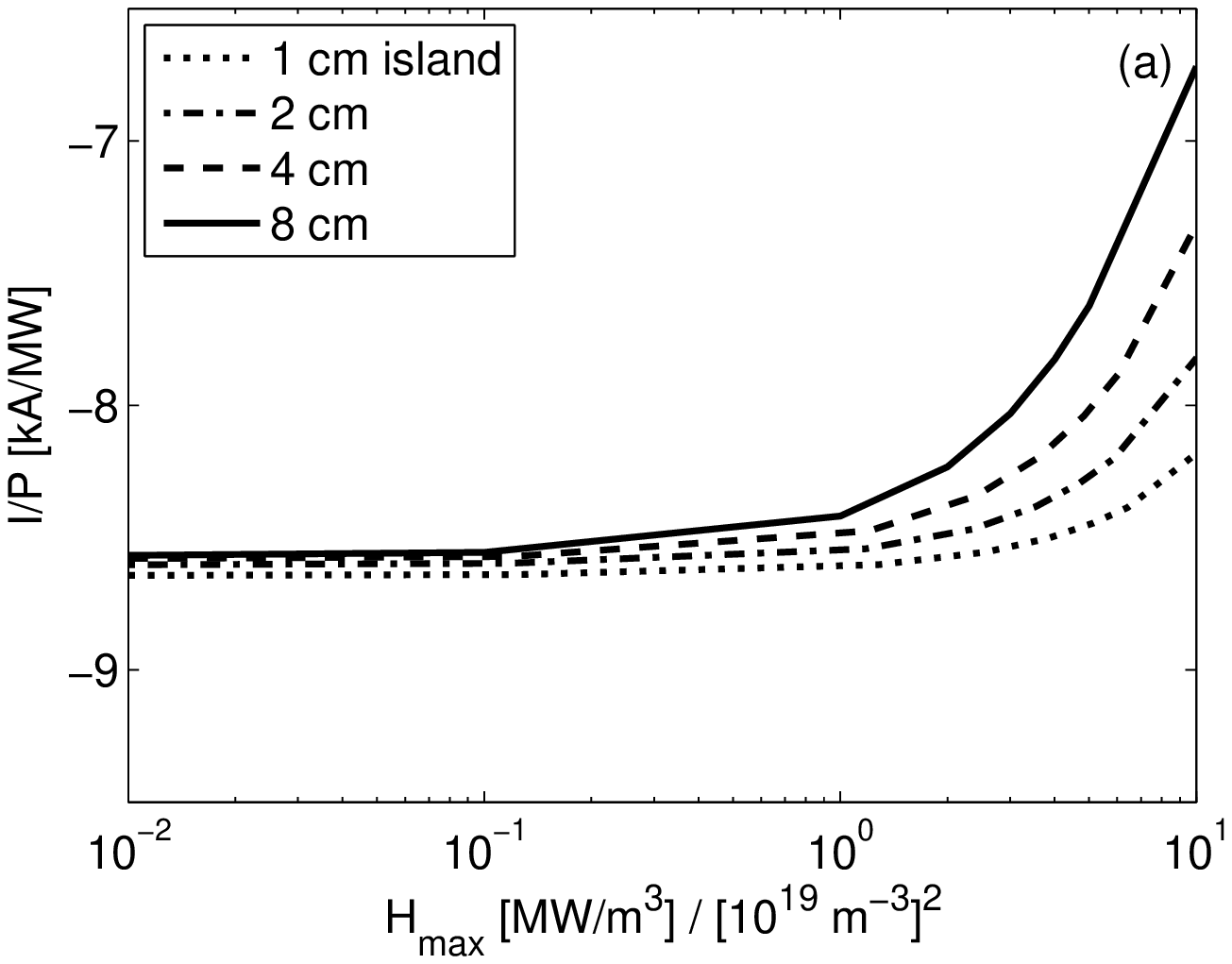}
\includegraphics[width=80mm]{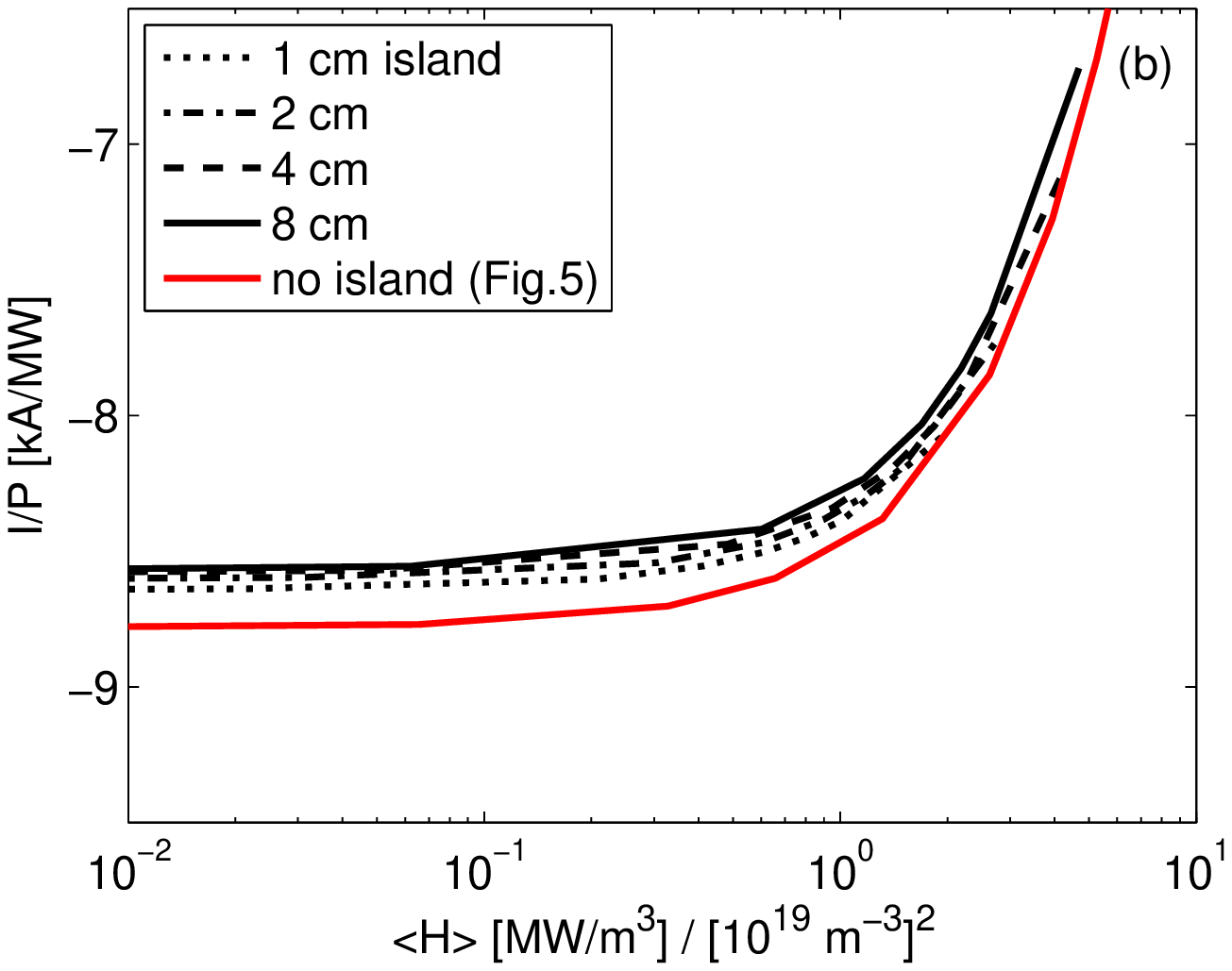}
\caption{\label{figure9}Global ECCD efficiency: (a) the global current drive efficiency $I/P$ as a function of the nonlinearity parameter $H_{\rm max}$ associated with the maximum in the power deposition profile; (b) the same but as a function of the profile averaged nonlinearity parameter $<H>$ as defined in Eq. (13). The different line styles refer to different island sizes as indicated in the legends. In red the result of Fig. 5 is repeated, which represents the results on the nonlinear current drive efficiency from a single ray, single volume shell RELAX calculation.}
\end{figure}

The non-linear effects introduced by exceeding the threshold have been studied in a series of RELAX calculations for injected powers in the range of $1$~kW to 10 MW. These calculations are performed for O-point heating and the results are summarized in Fig.~\ref{figure9}a, which depicts the global current drive efficiency as a function of the peak non-linearity parameter $H_{\rm max}$ for island sizes of 1, 2, 4, and 8 cm. As in the unperturbed equilibrium case, the non-linear effects are seen to result in a reduction of the absolute value of the current drive efficiency. However, the effect now appears to depend on the island size. This is because the relative volume of the region in which $H$ exceeds the threshold is a function of the island size. An averaged non-linearity parameter over the deposition profile can be defined as
\begin{equation}
\label{eq:averageharvey}
<H> \equiv {\int H p_{\rm ECCD} {\rm d}V  \over \int p_{\rm ECCD} {\rm  d}V}.
\end{equation}
When the results are plotted as a function of this averaged non-linearity parameter (as in Fig.~\ref{figure9}b), the data for all different island sizes overlap and coincide with those of calculations in case of the unperturbed equilibrium (see Fig.~\ref{figure5}).

\section{ECCD in the presence of a rotating magnetic island}
\begin{figure}[!hbtp]
\includegraphics[width=80mm]{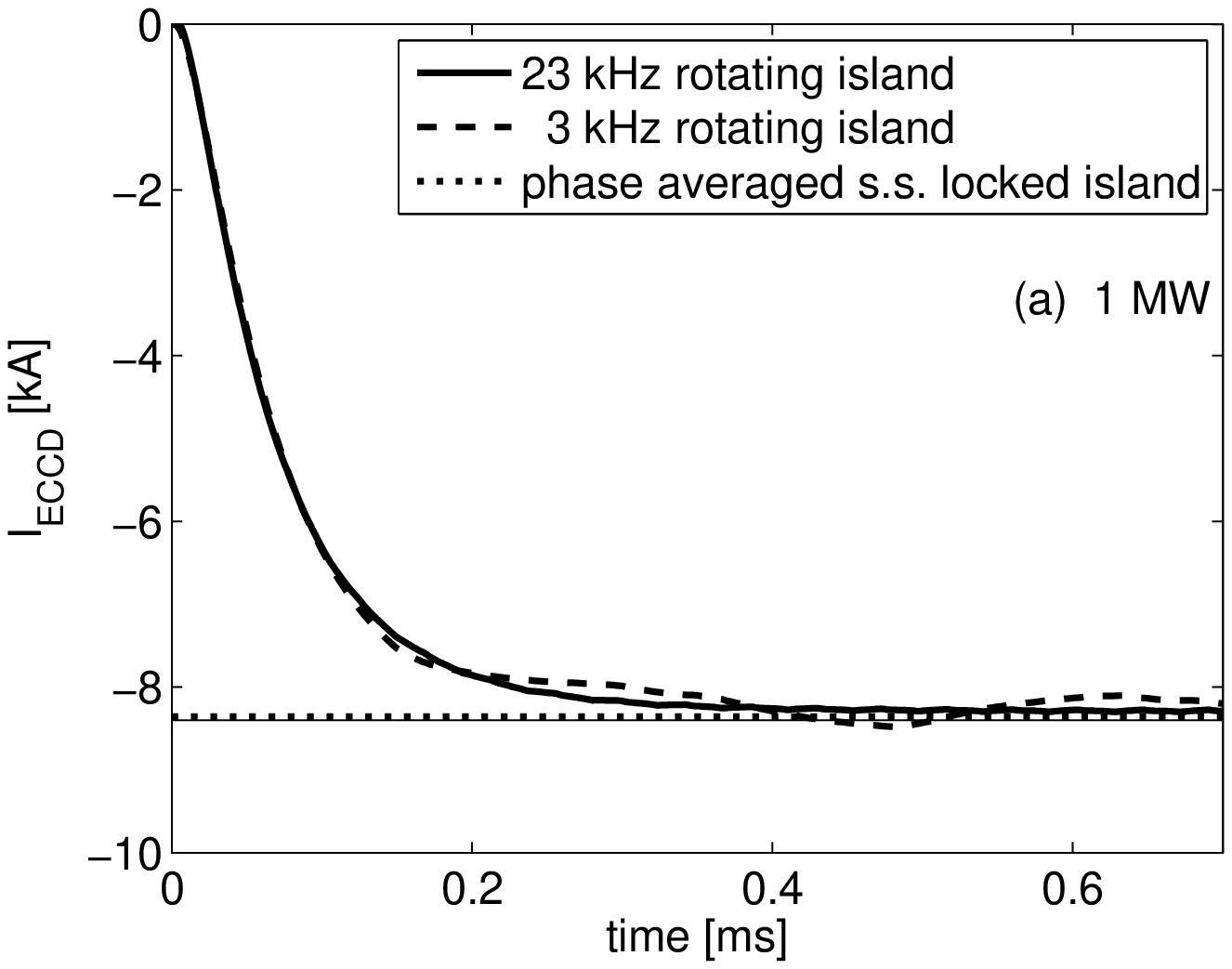}
\includegraphics[width=80mm]{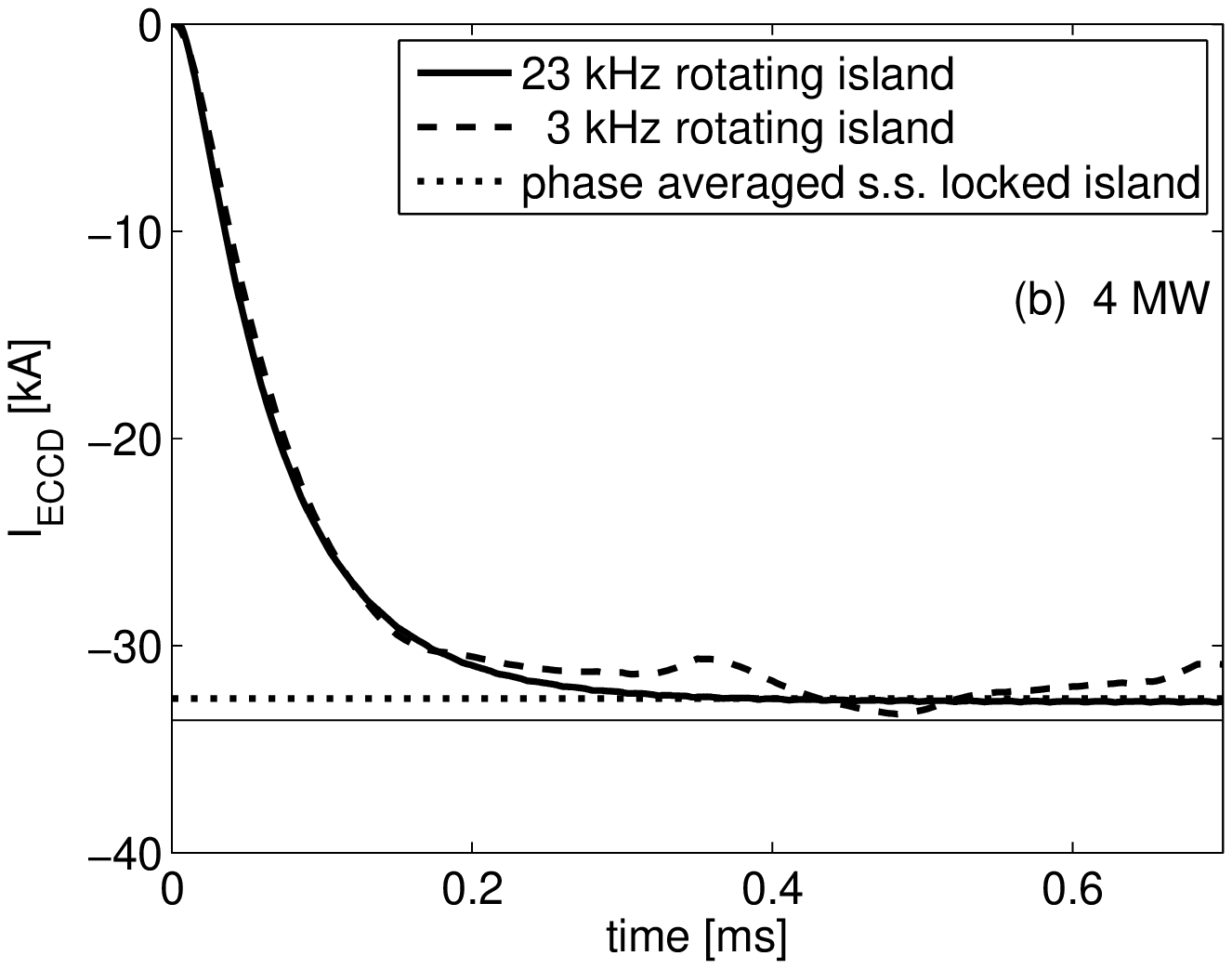}
\caption{\label{figure10}The total EC driven current as a function of time. Figure (a) presents the results for the injection of 1 MW and (b) those for 4 MW. The calculations are performed for an 8 cm wide 3/2 magnetic island rotating with a frequency of either 23 kHz (solid line) or 3 kHz (dashed line). The horizontal dotted lines refer to the result that is obtained by averaging the corresponding steady state driven current in a locked island over all different island phases. Similarly, the thin horizontal lines represent the phase averaged steady state current for a locked island extrapolated from a low power linear calculation.}
\end{figure}
We proceed with investigating ECCD when the island rotates. All ECCD simulations reported below, have been obtained for the fixed island size of $8$~cm. The injected power levels are 1~MW and 4~MW. The studied island rotation frequencies are $23$~kHz which is the actual value in the related experiment and $3$~kHz. The relevant collision frequency of the resonant electrons is about $1 \times 10^{4}$~Hz, i.e. between these two values. Note that these simulations were performed by successive calculations over short time intervals ($1/20$ times the rotation period) for $20$ discrete phases of the island. In all cases a total time interval of about $1$~ms was simulated which was sufficient to reach a quasi-steady state.

We remind that the $4$~MW power injection case corresponds to a power that is significantly higher than what was injected in the experiments. However, this case serves to illustrate the magnitude of the non-linear effects that can be expected in experiments at higher power or at lower density and consequently higher values of the nonlinearity parameter.

Figures~\ref{figure10}a and~\ref{figure10}b show the results in terms of the total driven current as a function of time for injected power levels of $1$~MW and of $4$~MW, respectively. For reference, each plot also contains a dotted line indicating the phase averaged steady state current in case of a locked island for the same injected power, as well as a thin solid line with the phase averaged steady state driven current extrapolating the low power linear results. In the $1$~MW power injection case the phase averaged locked island result is almost identical to the linear result, while the $4$~MW case shows a $3\%$ reduction in the absolute value of the current drive efficiency. The $3$~kHz oscillation is clearly visible in both figures, whereas the oscillation in the $23$~kHz cases is barely noticeable.

\begin{figure}[!hbtp]
\includegraphics[width=80mm]{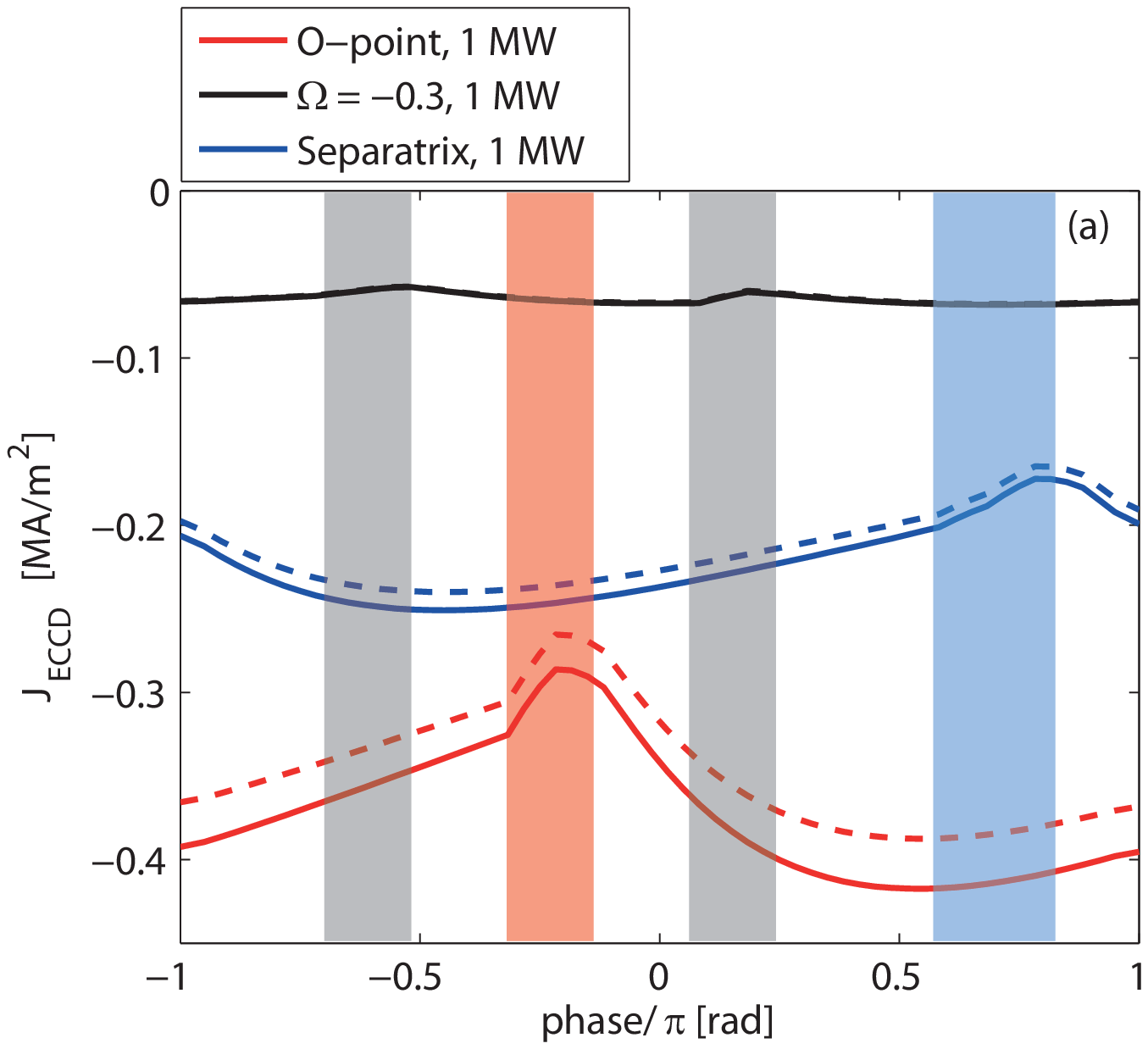}
\includegraphics[width=80mm]{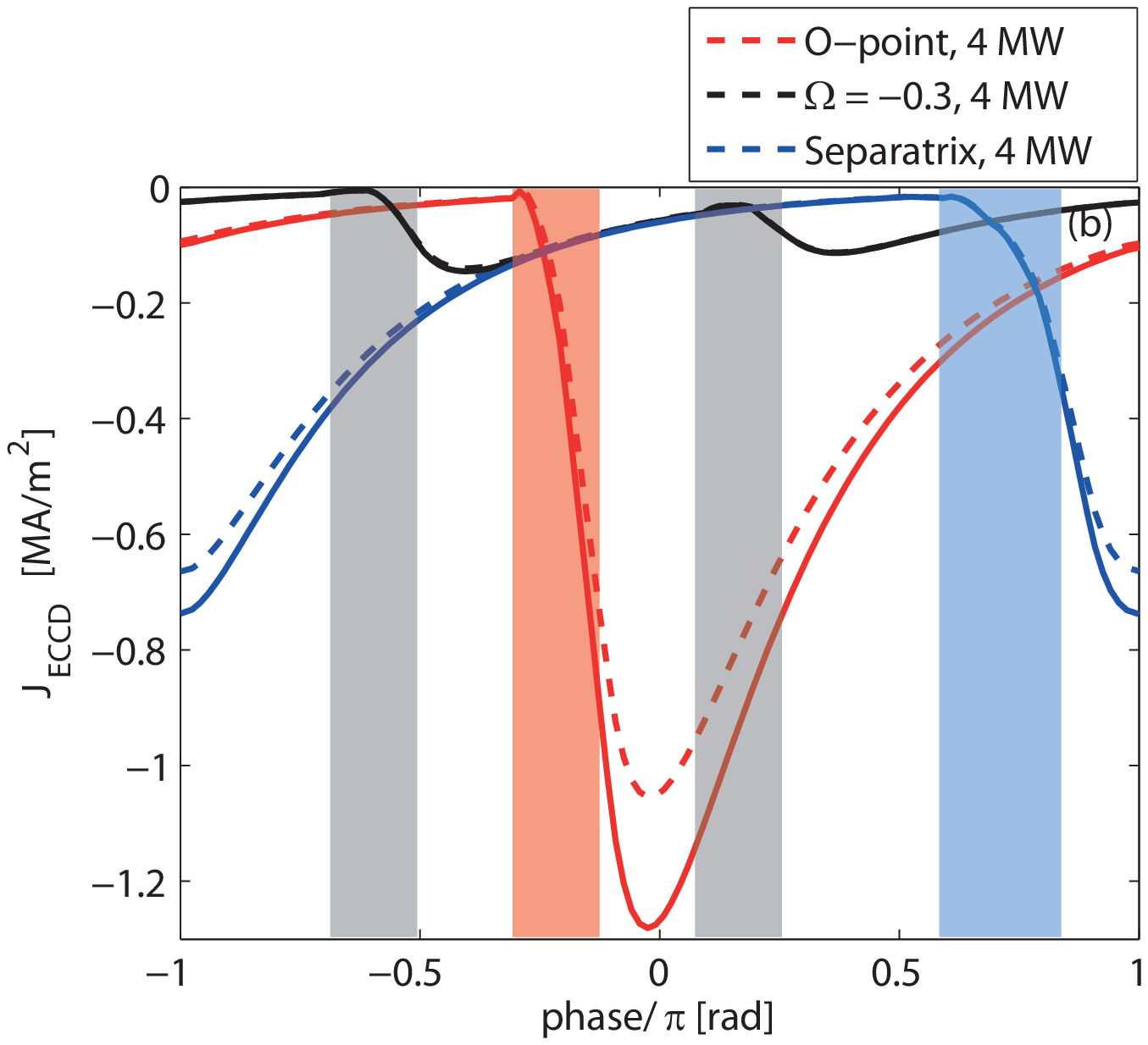}
\caption{\label{figure11}The driven current density as a function of the island phase, $\xi_0$ as defined in Section II.B. The results are plotted for an 8 cm wide 3/2 island rotating with either (a)  23 kHz, or (b) 3 kHz. The color of the curves refer to the position where the current density is given: red - for the maximum current density near the island O-point; black for the surface $\Omega = -0.3$ about midway the island; blue - for the surface immediate outside the outer separatrix. Full curves give the current density $J_{\rm ECCD}$ obtained with the injection of 1 MW, while dashed curves represent  $J_{\rm ECCD}/4$ as obtained with 4 MW to show that a fourfold increase in power results in the current increasing by less than a factor of 4, due to non-linearities. Shaded areas indicate the phase during which power is deposited on the surface of the corresponding color.}
\end{figure}

\begin{figure}[!hbtp]
\includegraphics[width=80mm]{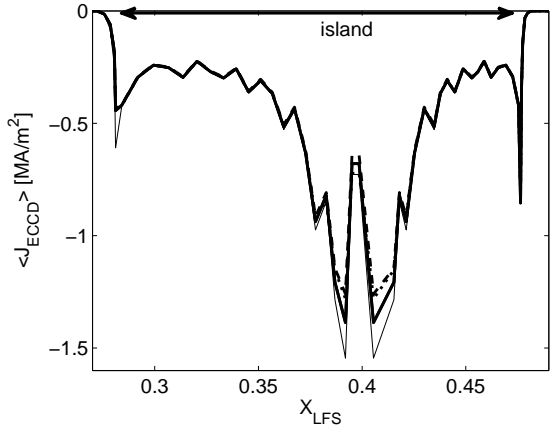}
\caption{\label{figure12}The averaged driven current density profiles for 4~MW power injection plotted as a function of the normalized low field side minor radius. The averaging is over a single rotation period during the quasi-steady regime in cases of a rotating island and over all phases for the locked island cases. The labeling of the different curves is the same as that of Fig.~\ref{figure10}b.}
\end{figure}

Figures~\ref{figure11}a and~\ref{figure11}b show the time evolution of the current density in a single rotation period during the quasi-steady state for an island rotating with $23$~kHz and $3$~kHz, respectively. The time is represented here in terms of the island phase $\xi_{0}$ as defined in Section II.B. The current densities shown are from the maximum near the O-point (red colored curves), a surface about midway the island ($\Omega = -0.3$, black colored curves), and the first surface outside the outer separatrix (blue colored curves). The shaded areas in the same colors indicate the times during which these surfaces are heated. Full curves refer to the results for 1~MW power injection while the dashed refer to those for 4~MW power injection. Note that the results for 4~MW are divided by $4$ in order to ease the comparison with the 1~MW results. We draw attention to a number of features that can be observed in these results. In particular, in the 23 kHz case (Fig.~\ref{figure11}a) a direct reduction in the driven current density (absolute value) is seen as soon as a surface is heated. This is a consequence of the Ohkawa effect resulting from the trapping of current carrying electrons~\cite{ohkawa}. This is an immediate result of the EC quasi-linear diffusion. The main current drive coming from the Fisch-Boozer effect is delayed by a collision time. As explained in~\cite{fisch,westerhof_eps2013}, the immediate effect of EC quasi-linear diffusion is mainly an increase of the perpendicular velocity of the resonant electrons. The reduced collisionality of these electrons subsequently results in the establishment of a current on a collisional time scale. As a result, the driven current density (absolute value) peaks well after the heating of the surface. Also note that the current drive efficiency both near the O-point and at the separatrix is significantly reduced in the 4~MW power injection case as compared to 1~MW power injection. In particular, for the lower rotation frequency of 3 kHz the maximum value of the efficiency reached near the O-point is reduced by $20\%$ in the 4 MW case. We remind that the driven current density at the O-point contributes most strongly to NTM stabilization.

Figure~\ref{figure12} shows averaged driven current density profiles for 4~MW power injection. The average is over a single rotation period during the quasi-steady regime in cases of a rotating island and over all phases for the locked island case. The results are plotted as a function of the normalized low field side minor radius.
The labeling is the same as that of Fig.~\ref{figure10}b. This figure shows that significant quasi-linear effects are limited to narrow regions around the O-point and immediately outside the separatrices. For the 1~MW injection case, the averaged current density profiles in all cases are practically identical to scaled low power locked island results.

\section{The consequences of non-linear effects in ECCD on the magnetic island evolution}
Having investigated the non-linear effects in ECCD extensively we now study their consequences on the magnetic island evolution. We conduct this study based on the generalized Rutherford equation (GRE)~\cite{rutherford1,rutherford2} written as:
\begin{equation}
0.82\frac{\tau _{\rm r}}{r_{\rm s}}\frac{\mathrm{d} w}{\mathrm{d} t} = r_{\rm s}\Delta ^{'}\left ( w \right ) + r_{\rm s}\sum_{i}\Delta ^{'}\left ( J_{\rm i} \right ),
\end{equation}
where $\tau_{\rm r}=\mu _0r_{\rm s}^2/\eta $ is the current diffusion time for the plasma resistivity $\eta$ at the resonant surface $r_s$ of the island. The first term on the right-hand side ${\Delta }'(w)$ is the classical stability index. The second term describes the modifications to the classical tearing mode equation as a consequence of all possible non-inductive contributions to the current perturbation. The term accounting for the stabilizing contribution of the EC driven current density $J_{\rm i} = J_{\rm CD}$~\cite{Hegna1997,sauter_2004,DeLazzari2009} is
\begin{equation}
\label{DeltaCD}
r_s\Delta '_{CD}=-\frac{16\mu _0 L_q r_s}{B_p \pi w^2}\left [\int_{-\infty}^{\infty}dx\oint d\xi J_{\rm CD}\cos \xi \right],
\end{equation}
where $x \equiv r - r_{\rm s}$ is the radial coordinate relative to the resonant surface, $\xi$ the helical angle in the cross-section of the island, $L_{\rm q}=q/{q}'$ the shear length, and $B_{\rm p}$ the poloidal magnetic field at the resonant surface. In the previous sections we have seen that the non-linear effects alter both the driven current density profile and the amplitude of the CD efficiency. We define a non-linear shape function $\Upsilon_{\rm CD}$ that accounts for all these effect in a single coefficient
\begin{equation}
\Upsilon_{\rm CD} \equiv \frac{1}{\eta _{\rm CD,lin}P}\int_{-\infty}^{\infty}\mathrm{d} x\oint \mathrm{d} \xi J_{\rm CD}\cos (\xi )
\end{equation}
where $\eta _{\rm CD,lin}$ represents the low power, linear CD efficiency calculated in a geometry in the absence of islands and $P$ denotes the total injected power in case of continuous wave (CW) application. Accordingly, the contribution of the current drive (Eq.~\ref{DeltaCD}) can be rewritten as,
\begin{equation}
\Delta_{\rm CD}^{\prime} = \frac{16\mu _{\rm 0}L_{\rm q}}{B_{\rm p}\pi}\frac{\eta _{\rm CD,lin}P}{w^{2}}\Upsilon_{\rm CD}.
\end{equation}
We present the consequences of the non-linear effects in ECCD on the magnetic island evolution in terms of this non-linear shape function $\Upsilon_{\rm CD}$.

\begin{figure}[!hbtp]
\includegraphics[width=80mm]{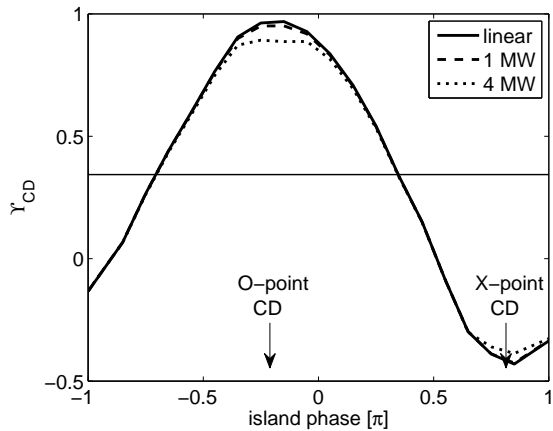}
\caption{\label{NTM_efficiency_locked}The non-linear shape function $\Upsilon_{\rm CD}$ as a function of the island phase $\xi_{\rm 0}$ plotted in the case a locked island of $8$ cm wide for increasing deposited power levels. The thin, straight line refers to the average of the low power, linear results over all phases of the island.}
\end{figure}

Figure~\ref{NTM_efficiency_locked} shows the non-linear shape function $\Upsilon_{\rm CD}$ as a function of the island phase $\xi_{\rm 0}$ in the case of a locked island of $8$ cm wide for increasing deposited power levels. The thin, straight line represents the average of the low power, linear results over all phases of the island. Although non-linear effects reduce the stabilization efficiency of O-point current drive as well as the destabilization efficiency for X-point current drive as the deposited power is increased, the difference is very small.

\begin{figure}[!hbtp]
\includegraphics[width=80mm]{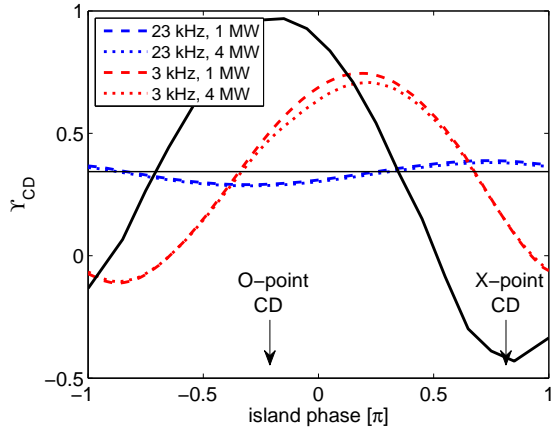}
\caption{\label{NTM_efficiency_rotating}The $\Upsilon_{\rm CD}$ profiles for a rotating island of $8$ cm wide. The blue curves represent the cases with an island rotation frequency of $23$~kHz. The dashed curves show the results for a deposited power of 1~MW and the dotted ones those for a deposited power of 4~MW. The red curves represent the cases with an island rotation of $3$~kHz. The black curve referring again to the low power, linear results in case of a locked mode, is included for comparison. The thin, straight line again represents the average of the linear regime results over all phases of the island.}
\end{figure}

Figure~\ref{NTM_efficiency_rotating} shows the $\Upsilon_{\rm CD}$ profiles in the case of a rotating island of $8$ cm wide. The blue curves represent the cases with an island rotation frequency of $23$~kHz. The dashed curves designate the results for a deposited power of 1~MW and the dotted ones those for a deposited power of 4~MW. The red curves represent the cases with an island rotation of $3$~kHz. The black curve representing again the low power, linear results in case of a locked mode, is included for comparison. The thin, straight line again represents the average of the linear regime results over all phases of the island. As in the case of locked islands, the consequences of the non-linear effects on the island evolution when the island rotation is taken into account is relatively small. In the 23 KHz case representing the actual experimental mode frequency, the effect is completely negligible even in the case where the power is increased to 4 MW. The figure does show another salient feature however: The phase shift of the $\Upsilon_{\rm CD}$ oscillation relative to the island rotation is clearly larger than what was observed in~\cite{ayten} on the basis of a simplified model for the dynamics of the EC driven current inside a magnetic island. This stems from the establishment of the Fisch-Boozer current on a collisional time scale as discussed in the previous section. Apart from this increase in the phase phase shift of the oscillation in $\Delta^\prime_{\rm ECCD}$, the effects of the island rotation obtained in the present work with the full Fokker-Planck modeling are consistent with the effects of island rotation as obtained in~\cite{ayten}.

\section{Summary and discussion}
In this paper we question to what extent the assumption of a linear ECCD efficiency is valid for the case of ECCD applied for the stabilization of NTMs inside a magnetic island. It has been shown by Harvey et al.~\cite{harvey} that non-linear effects appear when the ratio of the absorbed power density over the square of the electron density $n_{e}$ exceeds a certain threshold as represented by Eq.~\ref{eq:harvey}. Motivated by that, we calculate the non-linear ECCD efficiency through bounce-averaged, quasi-linear Fokker-Planck calculations in a realistic magnetic geometry including a locked or rotating magnetic island. For this purpose, we use the TORAY ray tracing code~\cite{Kritz,westerhof_toray} in combination with the RELAX bounce-averaged, quasi-linear Fokker-Planck code~\cite{westerhof_relax}. The unperturbed plasma equilibrium is taken in accordance with ASDEX Upgrade discharge nr. $26827$ which features a 3/2 magnetic island~\cite{reich}. We obtained the topology in the presence of a magnetic island by superimposing a helical flux perturbation $\tilde{\psi}$ which provides a realistic approximation to the radial dependence of the tearing perturbation as given in~\cite{Yang,hugovdbrand}. In this example the rays propagate tangential to the flux surfaces in the power deposition region. As a result, the whole power from a single ray is absorbed over a very narrow range of flux surfaces. These calculations show the possibility of significant non-linear effects on the ECCD efficiency, especially, in the case of locked islands or when the magnetic island rotation period is longer than the collisional time scale: because the volume enclosed by the flux surfaces inside the magnetic island in particular around the O-point is small, the local ECCD power density becomes very high and exceeds the threshold for nonlinear effects. The non-linear effects result in an overall reduction of the current drive efficiency for absorption of the EC power on the low field side of the electron cyclotron resonance layer in contrast with earlier claims~\cite{dasilvarosa}. This observation is explained by the quasi-linear flattening of the distribution function around the resonant velocities reducing the local absorption, which results in a further penetration of the power to regions in the plasma where the current drive efficiency is lower. As a consequence of the non-linear effects, also the stabilizing effect of the ECCD on the tearing mode will be reduced from the linear expectations. Nonlinear effects result in a global decrease of the ECCD efficiency. We present more detailed conclusions by dividing the results into three different categories. First we start with finding regarding ECCD in the presence of no island, that is in the unperturbed equilibrium.

\vskip 0.5 truecm

\textbf{ECCD in the unperturbed magnetic equilibrium}

\begin{itemize}
\item The maximum absorbed power density for the injection of a 1 MW ECW beam corresponds to a nonlinearity parameter $H_{\rm max} = 0.1 \left [ {\rm MW}/{\rm m}^{3} \right ]/\left [ 10^{19}{\rm m}^{-3} \right ]^2$. Thus, nonlinear effects are not expected for these conditions in ASDEX Upgrade in the absence of magnetic islands.

\item Low power absorption and driven current density obtained along the central ray of the full beam ray with both TORAY and RELAX show good agreement and therefore, provide a cross code benchmark between TORAY and RELAX in the low power, linear regime.

\item The threshold for non-linear effects is seen to be $H = 0.5 \left [ {\rm MW}/{\rm m}^{3} \right ]/\left [ 10^{19}{\rm m}^{-3} \right ]^2$ in agreement with Ref.~\cite{harvey}.

\item Nonlinear effects result in reduction of the absolute value of the global ECCD efficiency (i.e. the ratio of total driven current over total absorbed power). This reduction is explained by the fact that quasi-linear flattening of the distribution function reduces the absorption coefficient and shifts the peak in the $\rm{d} P/\rm{d} s$ profile to a smaller major radius, closer to the cold EC resonance, where the ECCD efficiency is lower.

\end{itemize}

\vskip 0.5 truecm

\textbf{ECCD in the presence of a locked magnetic island}

\begin{itemize}
\item The topology of the closed flux surfaces in the presence of islands leads to higher local absorbed power densities of the ECCD applied for magnetic island control.

\item The peak power density and, consequently, the likelihood of non-linear effects is strongly dependent on the phase of the island.

\item In case of deposition around the O-point of the island, the threshold for nonlinear effects is exceeded significantly for an injected power of 1 MW and for the parameters of the ASDEX Upgrade discharge being studied.

\item The significance of the non-linear effects is best parameterized by a profile averaged non-linearity parameter given by Eq.~\ref{eq:averageharvey}. As a function of this averaged non-linearity parameter, the non-linear ECCD efficiency is independent of island size and coincides with the results obtained in the unperturbed equilibrium.

\item Even at 4~MW the expected non-linear effects in the current drive efficiency do not exceed 10\% for the ASDEX Upgrade parameters used in this study.

\end{itemize}

\vskip 0.5 truecm

\textbf{ECCD in the presence of a rotating magnetic island}
\begin{itemize}
\item Rotation of the island generates an oscillation of the total driven current with the rotation frequency, which however becomes insignificant when the rotation frequency is larger than the collision frequency as is the case with a rotation frequency of 23 kHz.

\item For both 1 and 4 MW, the average of the total driven current over a rotation period converges to the same value as obtained by taking the phase average of the steady state driven current in the locked island cases. While for 1MW the time/phase averaged current is almost identical to the extrapolated linear result, the 4 MW case shows a reduction of the total driven current of about $3\%$.

\item The time evolution of the current density shows a direct reduction in the driven current density (absolute value) at the beginning of each period that a surface is heated. This is a consequence of the Ohkawa effect which stems from the trapping of current carrying electrons~\cite{ohkawa}. The main current drive arising from the Fisch-Boozer effect is delayed by a collision time~\cite{fisch,westerhof_eps2013}.

\item The current drive efficiency near the O-point and just outside the separatrices is considerably lowered when the power is increased from 1 to 4 MW. Especially, for the lower rotation frequency of 3 kHz the maximum value of the efficiency attained near the O-point is reduced by $20\%$ in the 4 MW case.

\item Time averaged driven current density profiles for 4~MW power injection shows that significant quasi-linear effects are constrained to narrow regions around the O-point and immediately outside the separatrices. For the 1~MW injection case, the averaged current density profiles in all cases are approximately identical to scaled low power locked island results.

\end{itemize}

We add that the non-linear effects may already be relevant at present power levels in case of locked islands whereas there is very little or no effect for present power levels when the island rotates. The consequences of the non-linear effects on the island evolution are found to be minor both in the case of locked and rotating islands.  At the mode rotation frequencies as observed in the actual experiments (about 23 kHz), the nonlinear effects are completely negligible even in the case where the power is increased to 4 MW.

\section{Appendix}
In the case of the ASDEX Upgrade discharge 26827 studied in this paper, the EC waves propagate almost tangentially to the flux surfaces of the unperturbed equilibrium in the region of EC power deposition. The original algorithm for the quasi-linear diffusion calculation in the RELAX bounce averaged Fokker-Planck code was based on the wave parameters in the point where the ray crosses the magnetic surface on which the distribution function is to be evaluated. In the case of a tangent ray this algorithm breaks down, because all power may in fact be absorbed in between the crossing of two of the discrete surfaces being considered. For this reason a new algorithm for the calculation of the quasi-linear diffusion operator has been implemented.

The surfaces ($i = 1, ..., n_{\rm surf}$) in the RELAX code represent finite volume shells $\Delta V_i(\psi_{\rm p,min,i} : \psi_{\rm p,max,i})$ in the toroidally symmetric unperturbed equilibrium or $\Delta V_i(\psi_{\rm h,min,i} : \psi_{\rm h,max,i})$ in the perturbed magnetic topology with magnetic island and each ray is discretized in small segments. Where $\psi_{\rm p/h}$ denotes the unperturbed poloidal flux $/$ perturbed helical flux. In the TORAY-FOM ray-tracing code, the EC wave beam is discretized  in a number of rays ($j = 1, ..., n_{\rm ray}$). A post processor divides the trajectory of each ray in small, subsequent segments $(k = 1, ..., n_{\rm seg})$ that lie entirely within one of the finite volume shells. For each of these segments the following information is extracted from the ray tracing results and transferred to the RELAX Fokker-Planck code:

\vskip 0.5 truecm

\begin{center}
\begin{tabular}{|l|l|}
  \hline
  variable & description \\
  \hline
  {\tt ENPAR} & the parallel refractive index $N_\parallel$ \\
  {\tt DNPARP} & the spread in $N_\parallel$ in the poloidal direction \\
  {\tt DNPART} & the spread in $N_\parallel$ in the toridal direction \\
  {\tt OMCOM} & the value of $\omega / \omega{ce}$ at the center of the segment \\
  {\tt DOMCOM} & the spread in $\omega / \omega{ce}$ over the segment \\
  {\tt OPTISC} & the length of the ray segment in cm \\
  {\tt BBO} & the ratio of the magnetic field $B$ at the center of the segment \\ & over the minimum field $B_{\rm min}$ along the flux surface \\
  {\tt NSCROS} & the number of the volume shell in which the segment lies \\
  \hline
\end{tabular}
\end{center}

\vskip 0.5 truecm

\noindent
The diffusion operator for volume shell $i$ is now obtained by summing the contributions from all segments of all rays within this volume shell:
\begin{equation}
    D_i = \sum_{j = 1}^{n_{\rm ray}} \sum_{k = 1}^{n_{\rm seg}} \delta_{{\tt NCROS(j,k)},i} D_{j,k}
\end{equation}
where each $D_{j,k}$ is evaluated according to Eq. (5).

\section{Acknowledgment}
\noindent The work in this paper has been performed in the framework of the NWO-RFBR Centre of Excellence (grant
047.018.002) on Fusion Physics and Technology. This work, supported by the European Communities under the contract of
Association between EURATOM/FOM, was carried out within the framework of the European Fusion Programme. The views and
opinions expressed herein do not necessarily reflect those of the European Commission.

\end{document}